# Diagramming the Class Diagram: Toward a Unified Modeling Methodology

Sabah Al-Fedaghi
Computer Engineering Department
Kuwait University
Kuwait
sabah.alfedaghi@ku.edu.kw

*Abstract*— **The object-oriented class is, in general, the most utilized element in programming and modeling. It is employed throughout the software development process, from early domain analysis phases to later maintenance phases. A class diagram typically uses elements of graph theory, e.g., boxes, ovals, lines. Many researchers have examined the class diagram *layout* from different perspectives, including visibility, juxtaposability, and aesthetics. While software systems can be incredibly complex, class diagrams represent a very broad picture of the system as a whole. The key to understanding of such complexity is use of tools such as diagrams at various levels of representation. This paper develops a more elaborate diagrammatic description of the class diagram that includes flows of attributes, thus providing a basic representation for specifying behavior and control instead of merely listing methods.**

## I. INTRODUCTION

Programming languages have progressed considerably to provide greater support for modeling and for abstract data types such as classes. In particular, object-oriented programming (OOP) has emerged as a preeminent mode of development in which a program is viewed as a set of interacting *objects* that consist of attributes and functions (methods). This approach has raised philosophical issues and led to insights regarding classes, objects, and other notions related to representation and modeling of reality. *Objects* have been singled out as an important concept [1]. According to Joque [1],

The history and philosophy that surround object-oriented programming offer a nuanced understanding of objects, their ability to hide part of themselves from the world, their relations, and their representation in languages… The philosophies that underlie OOP, likely as a result of the exigencies of creating functional systems, stress the relations between objects and the difficulties in conceptualizing objects as fully autonomous outside of the languages that address them.

UML (Universal Modeling Language) is standard for specifying, visualizing, constructing, and documenting the constructs of OOP. UML concepts are all well-known and applied in standards of software design. We assume that the reader is familiar with most of these concepts and their notations. In this work, we consider primarily *classes* in UML, as they are, in general, the most utilized elements in programming and modeling. A class diagram is "the heart and soul of an object-oriented design" [2].

Class diagrams are static structures that provide an overview of the system by specifying classes and the relationships between them. They are used for a variety of purposes such as understanding requirements, modeling the domain-specific data structure, and describing detailed design of the target system. "The class diagram is particularly useful through entire software development process, from early domain analysis stages to later maintenance stages" [3].

In object-oriented programming, it is typically claimed that a class in a computer program serves as a template for the creation of an object, just as Plato's forms were abstract philosophical templates for real world objects.

The notion of 'class' in object-oriented programming is Platonic to the extent that classes pre-exist objects in terms of program execution (as the Forms pre-exist material singulars), and that classes are used 'as a template for generating objects'. [4].

From the programming point of view, a class is effectively a pattern from which objects are created and defined in terms of *attributes* and *methods* (operations that the class can execute). Each attribute is described by name, type, and unique identifier; each method is described by name and a set of parameters. [5]. Objects encapsulate state and behavior, where a state is an instance of an attribute, and behavior is specified in the methods. The execution of a method is triggered by a *message*.

A class diagram is typically described using the boxes, ovals, lines,… of graph theory. Associations, dependencies, and inheritance relations are drawn as edges. Classes are represented by boxes containing three parts: the name of the class, its attributes, and its methods. Most diagrams in the UML specification have no more than 10 classes and 10 relations [6].

A great deal of research has explored object-oriented design, including how to identify classes and relations between them, and especially how to focus on the relevant aspects of the software system to be modeled [6]. Many researchers have examined the class diagram *layout* from different perspectives, including visibility, juxtaposability, and aesthetics, which affects "the costs of communication and to minimize misunderstandings" [6]. "In particular, the spatial layout of UML diagrams plays a crucial role in fostering program understanding" [7]. Dwyer [8] suggested that three-dimensional representation conveys more information than two-dimensional representation and applied this principle to layout of three-dimensional UML.

According to Dwyer [8], software can be incredibly complex, and examination of a source code shows "the extreme detail and tells us little [about] the operation of the system as a whole." The key to understanding of such complexity is to use tools such as diagrams (e.g., in UML) at *various levels* of representation. This notion of levels of representation has already been used in class diagram through *packages* in which classes are grouped together into a sub-diagram that facilitates viewing the model at a coarser granularity.

In this paper, we *develop a more elaborate diagrammatic description of the class diagram*. We open the black box of the diagram to inspect the "attributes and methods" of internal structure and activities components that normally appear in additional diagrams of UML.

Take for example a single class diagram as shown in Fig. 1, with its corresponding declaration in English. The diagram hardly increases illumination and understanding beyond what is expressed in English. It replaces the terms Class, Attributes, and Methods with rectangles and the position of rectangles in vertical order is supposed to indicate which one is which. The English description seems closer to programming language (e.g., C++); at least it is not enclosed in a box. The claim in this paper is that the class diagram can be expanded diagrammatically by yet another diagram.

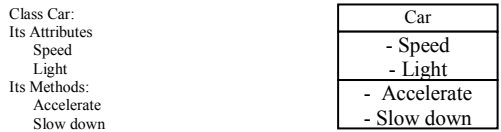

| Car |
| --- |
| - Speed |
| - Light |
| - Accelerate |
| - Slow down |

Class Car:
Its Attributes
Speed
Light
Its Methods:
Accelerate
Slow down

Fig. 1  Simplified class

Of course a *more elaborate diagrammatic description of the class diagram* can be accomplished by using other UML diagrams such as "replacing" names of methods with their corresponding activity or sequence diagrams. Eventually we can create a single unified diagram from the initial class diagram. This seems like a bad idea since no one has suggested such an approach; it is unprecedented.

UML specification defines two major *kinds* of diagrams: structure diagrams and behavior diagrams. The idea mentioned above (class diagram can be expanded diagrammatically by yet another diagram) leads to inserting behavior in the class diagram. Such a methodology that produces a *single-diagram* model is adopted by Dori [9] in Object-Process Methodology (OPM).

The OPM paradigm integrates the object-oriented, process-oriented, and state transition approaches into a single frame of reference. Structure and behavior coexist in the same OPM model without highlighting one at the expense of suppressing the other to enhance the comprehension of the system as a whole. [10] (Italics added)

This paper introduces a proposed conceptual methodology, called the *Flow(thing) Machines* (FM) model that produces a detailed single class description using a single diagrammatic language. A "machine" in this approach is a system component that creates, processes, and inputs and outputs *things* (to be defined later). In these approaches (i.e., OPM and FM), a single picture can be created that encompasses the details of the total system for purposes of understanding, design, and documentation.

Fig. 2 shows a general view of a single class of the proposed elaborated class diagram in comparison with the standard one shown to the right in Fig. 1. It involves two levels:

- A bottom part that embraces the static description in terms of attributes and flows of their instances;
- A top part comprising methods mapped to sub-graphs (enclosed in ovals) in the static description. A method is defined in terms of chronology of events (not shown in the figure) that specify the behavior of the class.

The figure seems complex, but it substitutes for many separate diagrams (e.g., activities, sequence, and state diagrams) while using a uniform language.

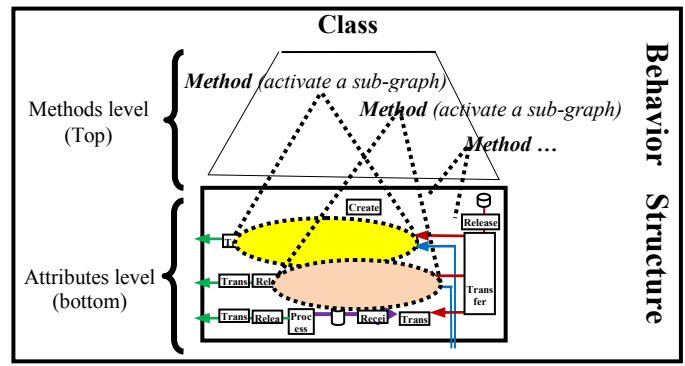

Fig. 2  FM representation of a class

## II. FLOWTHING MACHINES

To start modeling with the FM methodology, and for the sake of a self-contained paper, the next section briefly reviews FM, which forms the foundation of the theoretical development in this paper. The FM diagrammatic language has been adopted in several applications [11]–[16]; however, the example given here is a new contribution.

### A. FM model

"Flow things" are the "objects" of the FM model and include an almost limitless range of items, for example, passengers, luggage, signals, food, aircraft, data, attributes, and events, along with their dynamic behavior: i.e., flows. They flow in an (abstract) machine through five stages (states) in which they can be created, released, transferred, processed, and received (see Fig. 3). Hereafter, flow things may be referred to as *things* and an abstract machine as a *machine*.

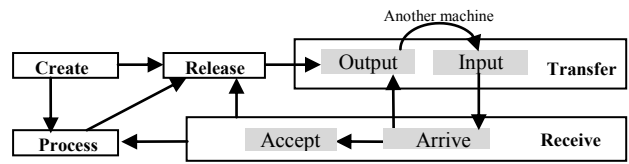

Fig. 3  Flowthing machine

The machine is the conceptual structure used to handle things as they pass through stages, from their inception or arrival to their de-creation or transmission out of the machine. Machines make up the organizational structure (blueprint) of any system. These machines can be embedded in a network of assemblies called spheres (e.g., an airport, a terminal) in which the machines operate. As will be shown later, all classes are spheres. For example, class *car* is a machine that includes sub-machines of engine, electrical system, and radiator, and it is the sphere that includes these sub-machines.

The machine shown in Fig. 3 is a generalization of the typical input-process-output model used in many scientific fields. The stages named in Fig. 3 can be described as follows:

**Arrive**: A thing reaches a new machine,

**Accept**: A thing is permitted to enter, or not. If arriving things are always accepted, Arrive and Accept can be combined as a **Receive** stage.

**Process** (change): A thing goes through some kind of transformation that changes it without creating a new thing.

**Release:** A thing is marked as ready to be transferred outside the machine.

**Transfer**: A thing is transported somewhere from or to outside the machine.

**Create**: A new thing appears in a machine.

The stages in FM are mutually exclusive. An additional stage of *Storage* can also be added to any machine to represent the storage of things; however, storage is not an exclusive stage because there can be stored processed things, stored created things, etc. The notion of spheres and subspheres refers to network environments. Multiple machines can exist in a sphere if needed. The machine is a subsphere that embodies the flow; it itself has no subspheres.

Triggering is the activation of a flow, denoted by a dashed arrow. It is a dependency among flows and parts of flows. A flow is said to be triggered if it is created or activated by another flow. Triggering can also be used to initiate events such as starting up a machine.

### B. Example

According to Maciaszek [2],

The use case model is the main UML representative and the focal point of behavior modeling… In practice, the importance of use cases goes even further. Use cases drive the entire software development lifecycle, from requirements analysis to testing and maintenance. They are the focal point and reference for most development activities… It is worthwhile emphasizing…, that a use case model can be viewed as a generic technique for describing all business processes, not just information system processes.

Maciaszek [2] gives an example of a video store transaction, describes its use case (see Fig. 4) and requirements as follows:

The Customer decides to pay for the video rental and offers cash or debit/credit card payment. The Employee requests the system to display the rental charge together with basic customer and video details. If the Customer offers cash payment, the Employee handles the cash, confirms to the system that the payment has been received and asks the system to record the payment as made. If the Customer offers debit/credit card payment, the Employee swipes the card, requests the Customer to type the card's PIN, select debit or credit account and transmit the payment. Once the payment has been confirmed electronically by the card provider, the system records the payment as made.

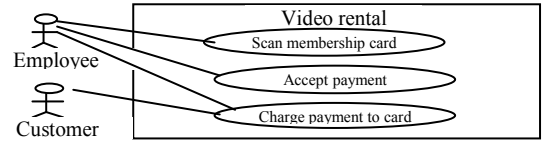

Fig. 4 Use case diagram (redrawn, partial from [2])

### B.1 Static FM description

Fig. 5 shows the FM static representation of this example with some minor modification. Assuming that the customer brings chosen videos to the employee, the employee signals the system to display the sale page (circle 1) to be processed (2). The employee then inputs data for each video (3) into the system (4), where the data are recorded and the price is calculated. Note that this is a repeating process (5). Repetition is a thing that can be created, processed, … (6). The end of the process triggers (7) creation of the total price (8).

This is followed by selection of payment type (9); if the payment is by credit card, this triggers an instruction to insert the card (10). The employee inserts the customer's card (11), which is processed by the system (12). Note that the card "embeds" its number (13), thus "transferring the card" implicitly implies transferring the number. If the card is OK, an instruction is sent to input the PIN number (14). Accordingly, the customer inputs the PIN number (15). Using the PIN number (16), the card number (13), and the total payment (8), the program creates a request for payment (17) and sends it to the paying agency (18). A receipt is received (19), indicating approval of the payment, and sent to the customer (20) – or a negative response is received (21). Note that Fig. 5 can easily be extended to model further details and exception cases such as requesting the insertion of a different card in case of rejection.

### B.2 Behavior description

According to Maciaszek [2], "System behavior is what a system does when it is responding to external events. In UML, the outwardly visible and testable system behavior is captured in use cases. Consistently with a model being able to be applied at various levels of abstraction, a use case model may capture the behavior of a system as a whole or the behavior of any part of the system – a subsystem, component or class."

In FM, *behavior* involves the flow of things during *events* *w*hen the static description (e.g., Fig. 5) is acted upon. The chronology of the resultant activities can be identified by orchestrating the sequence of these events in their interacting processes (machines).

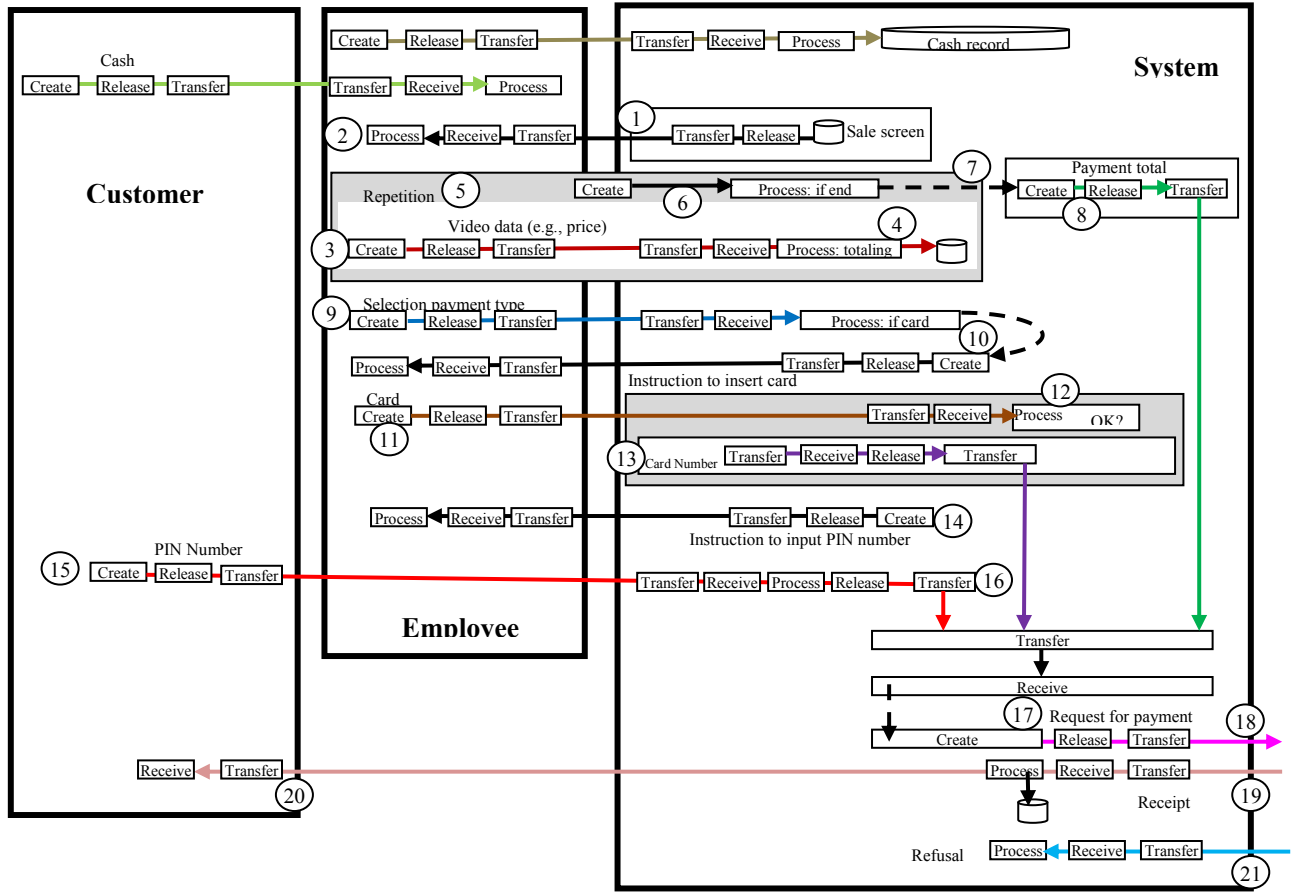

Fig. 5  FM representation of video rental use-case example

In FM, *an event is a thing like other things: i.e., what can be created, processed, released, transferred, and received*. An event *activates* flows (sub-diagrams of the description shown in Fig. 5). Note that the *process* stage of an event means that an event *runs its course*. Accordingly, the choreography of execution can emerge from the arrangement of events.

In FM, modeling proceeds as follows:

*Static description → Event-ized specification → Control*

Modeling of behavior takes place in a phase that occurs after the static description is created (e.g., Fig. 5) and involves modeling the "events space," where an event takes place or happens.

An event is specified by its spatial area or subgraph, its time, the event's own stages, and possibly by other things, e.g., descriptions such as intensity or extent (strength). Note that a *conceptual* event refers to sets of (elementary) events extended in space and time that, in the context of the involved model, together form a meaningful event. Events appear at different levels, e.g., creating a thing is an event in itself; however, modeling focuses on "meaningful" events. Not every event in history is a historic event. Fig. 6 shows the event: *A customer pays in cash for the rented videos*. It includes time of event, the occurrence of the event and its duration, and the "event space" as a sub-diagram of Fig. 5.

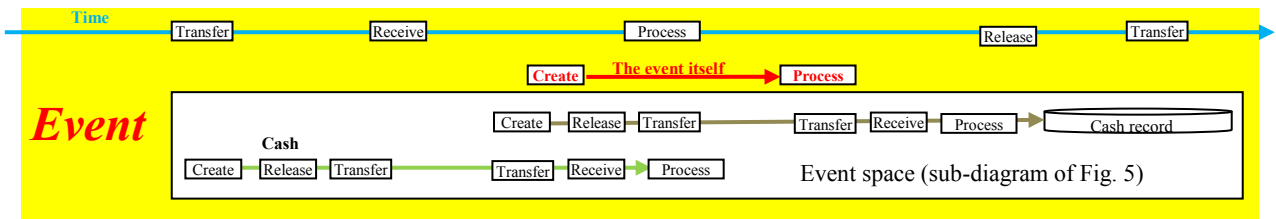

Fig. 6  FM representation of an event

To simplify the drawings of events, here we will not draw the time and stages of the events but focus on their spaces. We can select events of interest in the video rental system shown in Fig. 7 as the following seven events:

- Event 1 ($V_1$): Displaying the sale screen (assuming it is initialized) and
- Event 2 ($V_2$): Entering data information for the rented videos.
- Event 3 ($V_3$): Selecting payment by card and inserting it.
- Event 4 ($V_4$): Inputting the PIN number and constructing a request for payment that is sent to the payment agency.
- Event 5 ($V_5$): Approval of payment and sending a receipt to the customer.
- Event 6 ($V_6$): Refusal of payment.
- Event 7 ($V_7$): Payment by cash.

Accordingly, a control for the system can be developed by identifying the permitted chronology of events, as shown in Fig. 8. The brackets around $\{V_2\}$ indicate repeating the same event for each selected tape.

## III. APPLYING FM TO CLASSES

In FM, a class includes a single class machine and several "attribute" machines. The "methods" are processes in the class machine.

Deitel and Deitel [17] give an example that creates the class *Time* (as shown partially in Fig. 9) and a driver program that tests the class. Note that, in general, a class designation in UML is not (necessarily) the same as class in a programming language. Fig. 10 shows the general FM template of the flow with three methods in the given class. The flow arrows represent the life cycle of an object: an object is either created by a constructor or set in the main program, then processed and may be sent somewhere. Note that an object is a *thing* that can be created, processed, ….

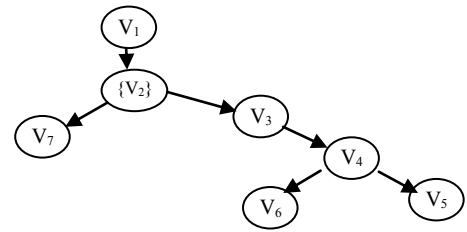

Fig. 8 Defining the chronology of events

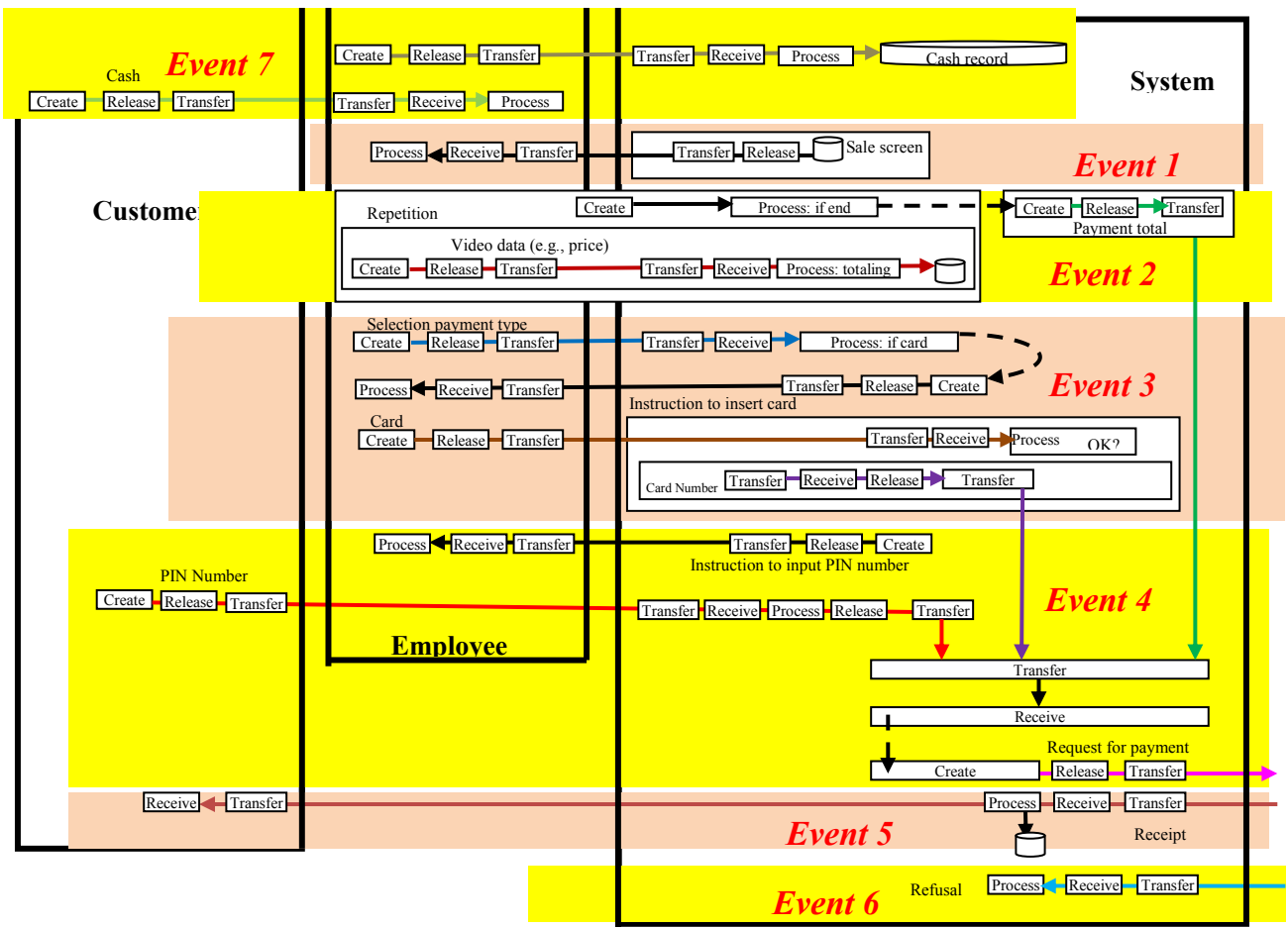

Fig. 7 FM representation of the example

```
class Time {
public:
Time();
void setTime( int, int, int );
void printUniversal();
void printStandard()
private:
int hour; int minute;int second; };
Time::Time()
{ hour = minute = second = 0; }
void Time::setTime( int h, int m, int s ) {
hour = ...;  minute = ...;  second = ...; }
void Time::printUniversal(){
cout ... << hour << ... << minute ...<< second;}
```

Fig. 9  Partial view of some statements in class *Time*

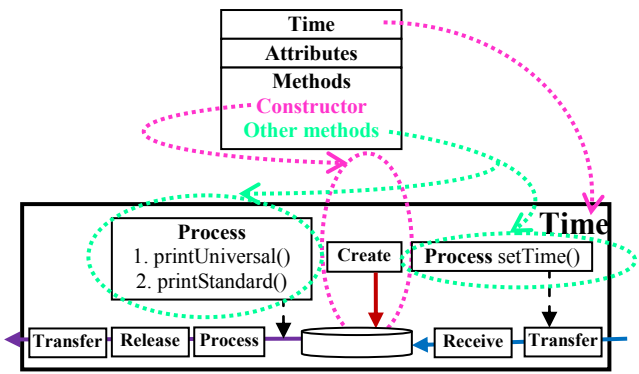

Fig. 10  FM representation of class *Time* and mapping of elements of classical diagrammatic description except attributes

The attributes define this object structure. In Deitel and Deitel's [17] example, they are expressed as shown in Fig. 11, where methods are included for illustration purposes since methods will appear at the behavior level, as will be described. The attribute values are either initialized to 0.0 (instructor) or set from the outside (main).

In Fig. 11, the class *Time* structure includes three attributes: *hour, minute*, and *second*, and three methods: *setTime, printUniversal*, and *printStandard*. In the figure, class *Time* is created (circle 1), causing triggering of the constructor that inserts zeros (2) in *hour* (3), *minute* (4), and *second* (5). The process (method) *setTime* (6) inserts (7) new values in *hour*, *minute*, and *second*. The processes *printUniversal* and *printStandard* cause the processing of *hour*, *minute*, and *second* according to the designated process and sends them to the outside (8, 9, and 10; e.g., C++ *cout*).

The methods can be conceptualized as events laid over the static description. Assuming the statements,

*Time t;*
*t.setTime( 13, 27, 6 );*
*t.printUniversal();*
*t.printStandard();*

Events fall into four types, as shown in Fig. 12, where *t.printUniversal();* and *t.printStandard();* are represented in one sub-diagram since the difference between them consists of details in object processing. Accordingly, the operational semantics of the statements is the execution of events e1, e2, e3, and e4, in that order.

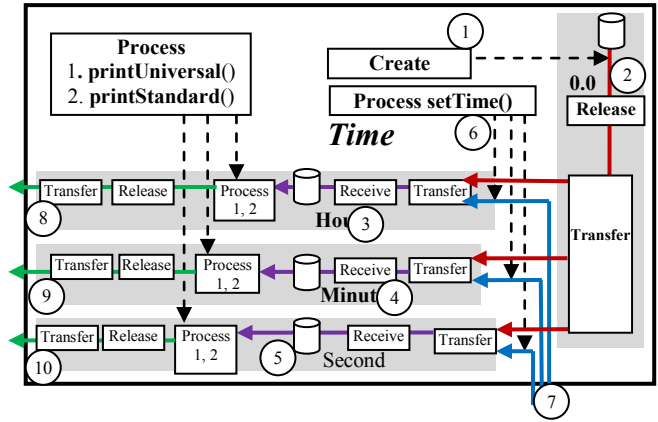

Fig. 11  FM representation of class *Time*

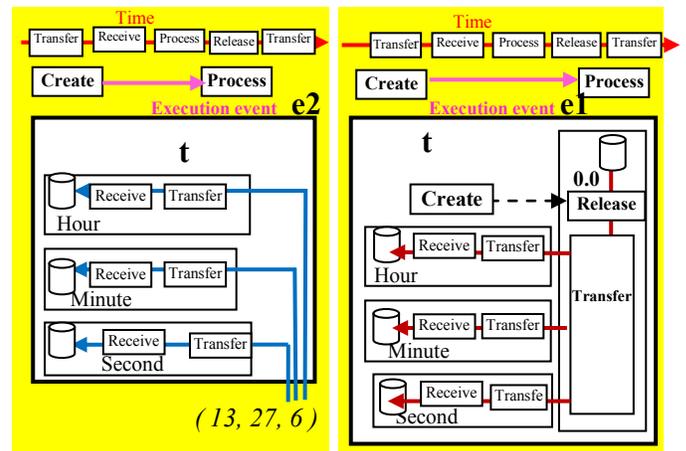

(a) setTime()       (b) Initialization to zeros

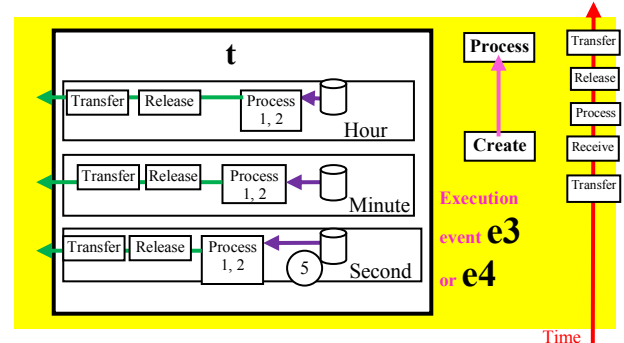

(c) printUniversal() and printStandard()

Fig. 12  FM representation of types of execution events

Fig. 13 shows the two levels: structure and behavior, as specified in the introduction in Fig. 1, where only the event space for *t.printUniversal();* and *t.printStandard();* is shown. This two-level conceptualization should be emphasized for conceptualizing class in object-oriented modeling. The standard practice of defining classes diagrammatically can cause misunderstanding, e.g., for students. Instead, we propose the diagram shown in Fig. 14.

For so-called *object diagram*s, Fig. 15 shows the diagram of a certain object through actual time (heavy arrows, not to be confused with class *Time*), where the methods play a role similar to events in a state machine diagram.

## IV. CLASS WITH INHERITANCE

Wikibooks [18] gives the class definition *car* that is shown partially in Fig. 16. Fig. 17 shows its FM representation. Since the initial constructor is illustrated in the previous example and for the sake of simplicity, it will not be included in the current example. In the figure, the car (circle 1) includes five processes: *setSpeed* (2), *getSpeed* (3), *getFuel* (4), *drive* (5), and *refuel* (6) that perform their tasks as follows:

```
class car
private maxSpeed ...
public fuel ...
public sub setSpeed(...)
    maxSpeed = s
end sub
public function getSpeed() ....
    return maxSpeed
    end function
public sub refuel(...)
    console.writeline("pumping gas!")
```

```
fuel = fuel ÷ x
end sub
public function getFuel() ...
    return fuel
    end function
public sub drive()
    fuel = fuel - 1
end sub
end class
```

Fig. 16  Class *Car*

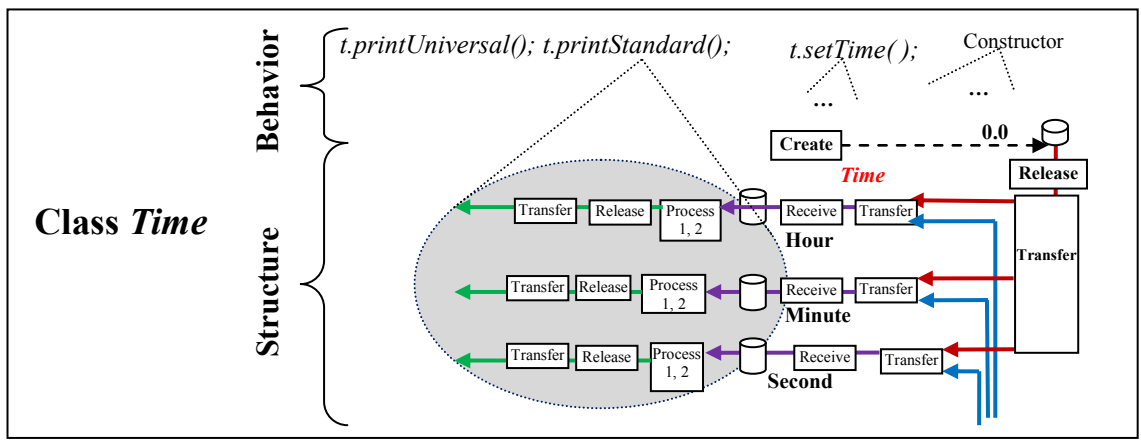

Fig. 13  General view of a single class as applied to the class *Time*

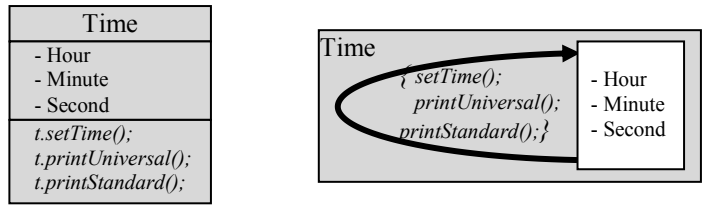

Fig. 14. Standard and proposed definitions of class

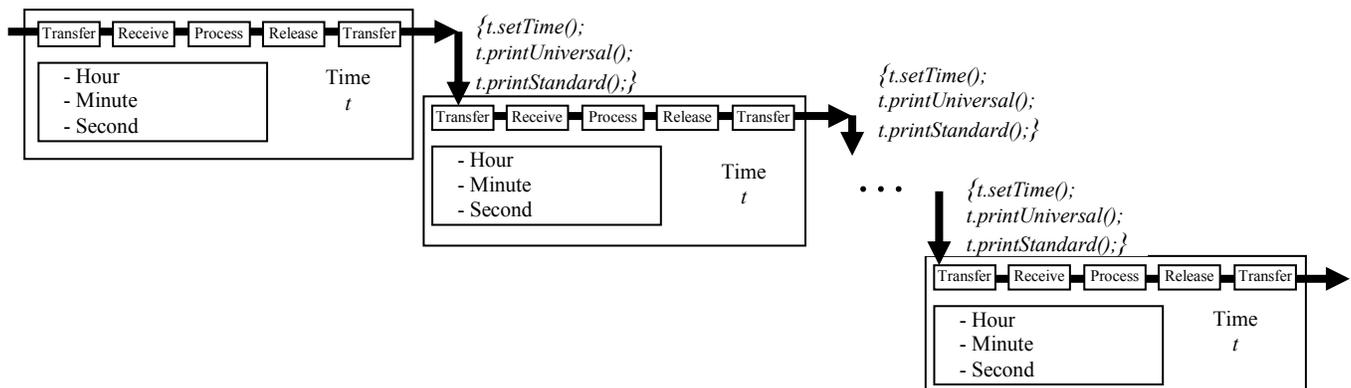

Fig. 15  FM representation of *object diagram* (heavy arrows represent actual time)

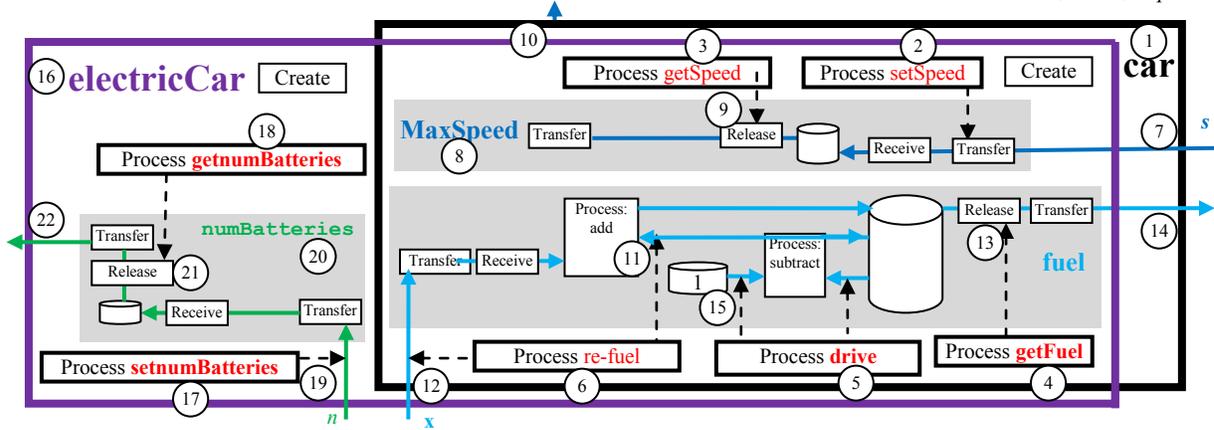

Fig. 17  FM representation of classes *Car* and *electricCar*

- *setSpeed* (2) triggers inputting *s* (7) in maxSpeed (8).
- *getSpeed* triggers (9) outputting (10) maxSpeed.
- *Refuel* (6) triggers adding fuel (11) to input x (12).
- *getFuel* (4) triggers (13) outputting (14) fuel.
- *Drive* (5) triggers subtracting 1 (15) from the fuel

Also given is sub-class *electricCar* (see Fig. 18), which inherits relevant attributes and methods in class *Car*. This class is also included in Fig. 17, where the class *electricCar* (16) has two additional processes (methods), *setnumBatteries* (17) and *getnumBatteries* (18), as follows:

- *setnumBatteries* (17) triggers inputting *n* (19) in numBatteries (20)
- *getnumBatteries* (18) triggers (21) outputting (22) maxSpeed

## V.  RELATIONSHIPS

Relationships between classes include generalization (an inheritance relationship) and association (dependency, aggregation, and composition). This section discusses two simple samples that embed relationships.

### A.  *Class book*

Hock-Chuan [19] gives the class diagram shown in Fig. 19 (some details have been omitted). It is assumed that a book is written by one and only one author. Fig. 20 shows a static FM representation of the class *Book* developed according to our conceptualization.

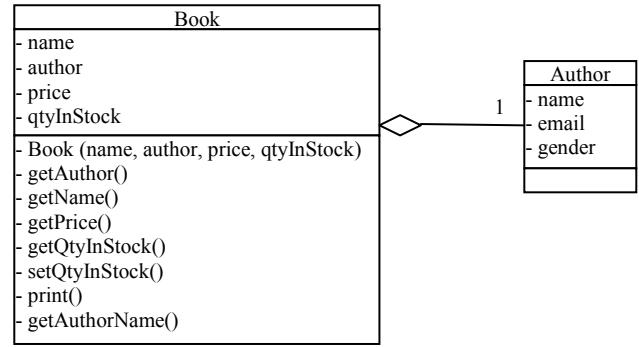

Fig. 18  Class *electricCar*

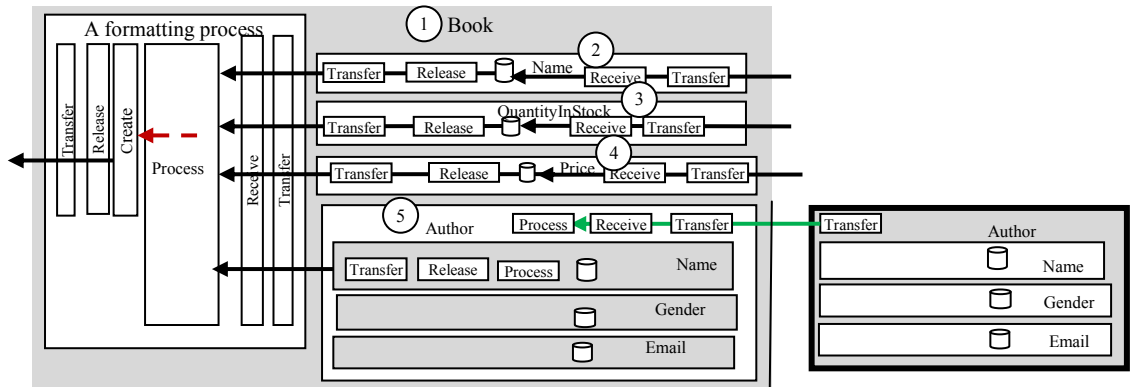

Fig. 19. Class *Book*

Fig. 20. Static FM representation of class *Book*.

The figure involves the flows of different attributes from the class (process *get*) where *Author's name* is imported from class *Author*. In the figure, the Book sphere (1) includes the attributes *Name*, *Price*, and *qtyInStock* (2-4). It also includes the sub-sphere *Author* (5). The relation that a book has only one author is embedded in the representation since there is only one author sub-sphere within the book sphere (class). The identification of methods occurs during the next phase when behavior is defined through the set of events.

Fig. 21 shows selected events in class *Book* as follows:

e1: *Receiving* object *Author* to construct object *Book*
e2: Processing and sending data (e.g., attribute, object of book, …) to output
e3: Processing *Author's name* and sending it to e2.
e4: Sending Price to e2
e5: Sending Name to e2
e6: Sending QtyInStock to e2
e7: Setting QtyInStock

Accordingly, the methods (compound events) can be defined in terms of events as follows:

getAuthor: $\{e_1\}$
getName: $\{e_5, e_2\}$
getPrice: $\{e_4, e_2\}$
getQtyInStock: $\{e_7, e_2\}$
setQtyInStock: $\{e_6\}$
getAutherName: $\{e_3\}$
Book (name, author, price, qtyInStock) : $\{e_5, e_1, e_4, e_7, e_2\}$

## B. *Automated Teller Machine System*

The site *Source Code Solutions* [20] presents a class diagram of an Automated Teller Machine (ATM) system, simplified and partially shown in Fig. 22. Fig. 23 shows the corresponding FM representation. The process starts with the client inserting a card (1) with a number that flows to the machine where it is processed (2). We assume here many simplifications such as an initial screen and no exceptions or errors occurring during operations. Processing the card triggers a new display (3). Here it is assumed the client then selects the withdrawal operation (4; the deposit operation selection will be described later).

It is also assumed that the required data are input and received (5). Accordingly, the data of the withdrawal operation are processed to create a transaction (6) that flows to the bank (7) where it is processed (8) to:

- trigger (9) updating of the account (10)
- Store the transaction (11).

A confirmation (12) is sent to the machine that flows to the client (13) causing it to,

- Display instructions (14)
- Release the card back to the client (15)
- Release cash to the client (16).

If the client selects deposit of cash (17), the request is processed (18). The client then inputs cash that flows to the machine (19) and is stored (20). A transaction is created (21) that flows to the bank (22) to be processed (23). The account is updated (24) and a confirmation (25) is sent to the machine (26) and to the client (27).

Note that some attributes such as name of the client and the machine are not included since they are not relevant to the processes, e.g., client's name is embedded in the card.

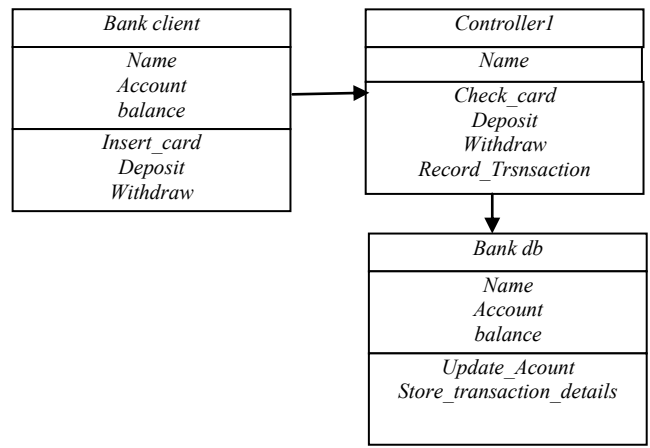

Fig. 22 Class diagram of ATM (redrawn, partial from [20])

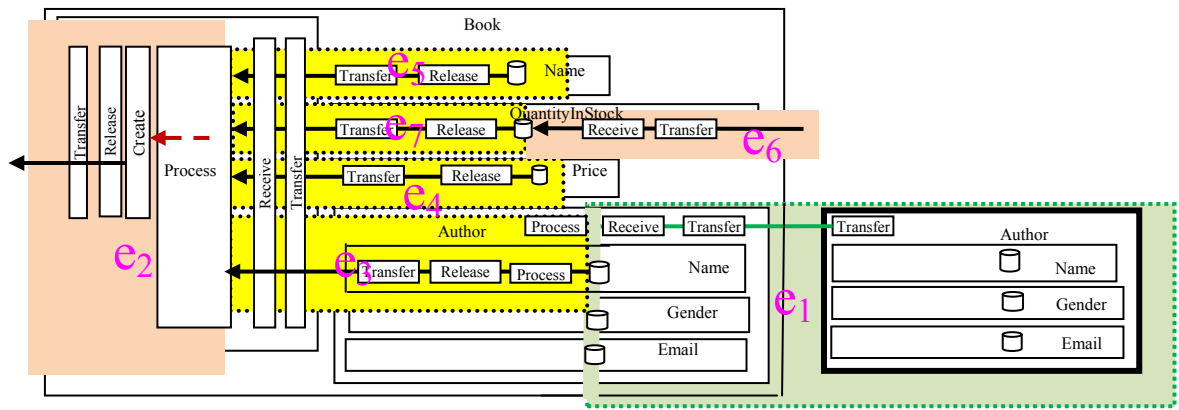

Fig. 21 Events in Fig. 20

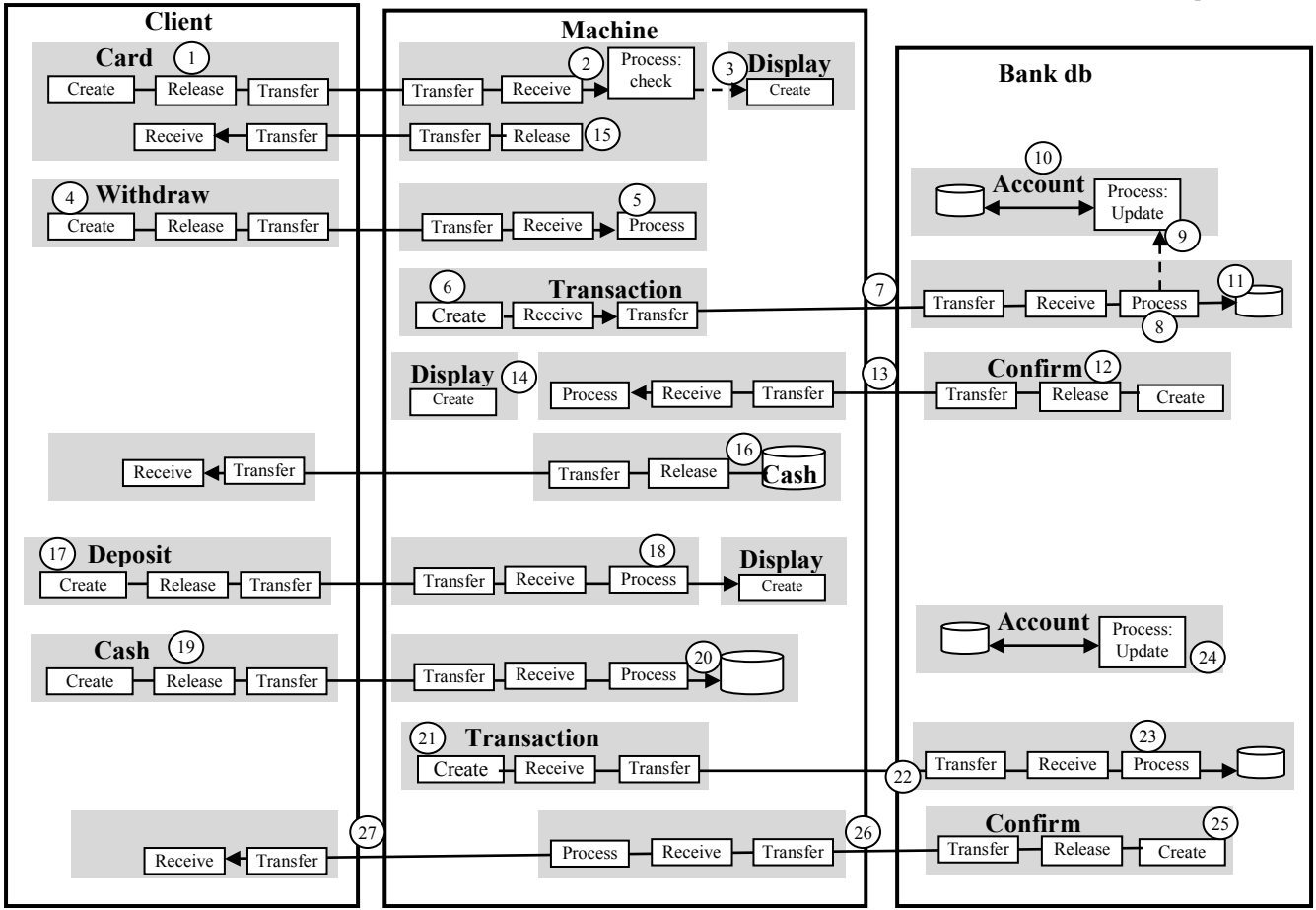

Fig. 23  FM representation of ATM

Because of space considerations, only the upper part of Fig. 23 is "event-ized," as shown in Fig. 24. This set of events can be utilized to develop a control module for the ATM system by building a sequence of event occurrences, as shown in Fig. 25.

### C.  Comments

It seems that a class is a type of sphere. The FM representation of classes emphasizes the difference between static description (attributes) and description of behavior (methods).

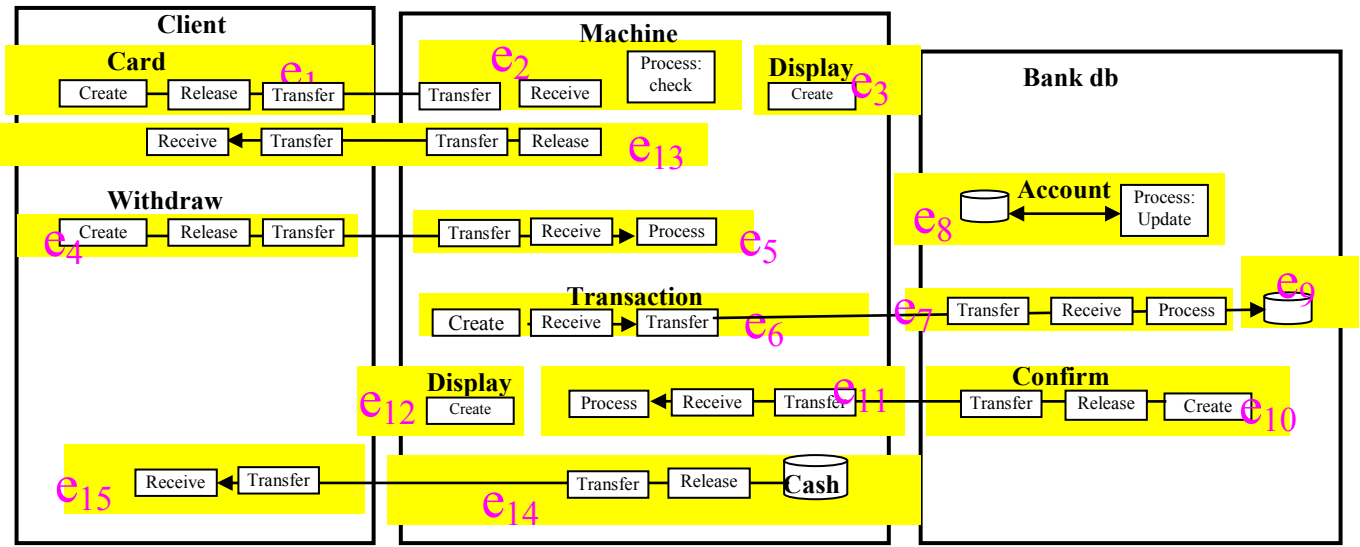

Fig. 24. Events in the ATM

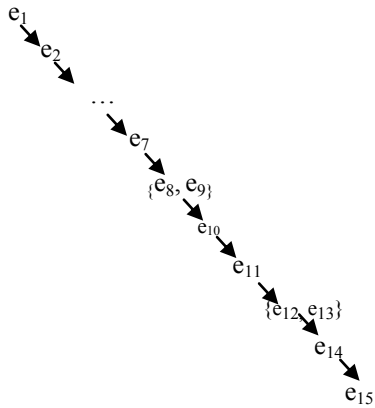

Fig. 25. Sequence of events

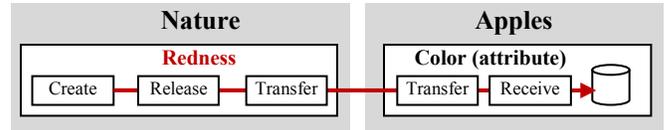

Fig. 27. Defining the attribute of color red by the basic thing Redness

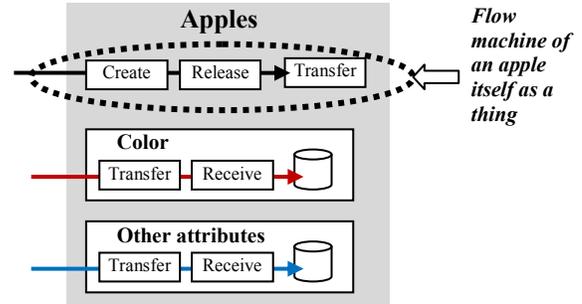

Fig. 28. The Apple as a sphere and a sub-sphere Redness

A static description is far richer since it maps the basic flows in the class. Relationships turn out to be mostly constraints on the flow. For example, receiving a card could be modelled to trigger blocking receipt of any additional cards, i.e., from only one client at a time; receiving cash in deposit could be repeated until the entered banknotes are exhausted,.

Note that basic types can be added to the FM representation, as exemplified in Fig. 26. A *Real number* is a basic thing. For example, consider the following,

> If two apples are both red, this is because there is a *Form of Red* that is able to manifest itself in both those apples at once. There is the individual, a particular apple (the thing); there is the red of that apple – which exists right "in" or with that apple; and finally, there is the *Form of Red*, which manifests itself in the red of this apple (and of course, the red of other apples). ([21], M. C. MacLeod and E. M. Rubenstein, "Universals", Internet Encyclopedia of Philosophy (no date). http://www.iep.utm.edu/u/universa.htm#H1) (Italics added)

The basic (unstructured and decomposed) attributes of Color *red* are defined as shown in Fig. 27. Such a conceptualization contrasts with current ontologies. For example in BWW-Ontology [22], properties such as "addresses" and "jobs" are not things. "The world is made up of substantial things that exist physically in the world" [23]. Here, it is not clear whether "color" is a thing. In FM, color is a *thing* since it can be created (appear, exist), be processed, …, and an apple is a thing since it can be created, processed, …, and the apple can be a sphere with attributes (see Fig. 28).

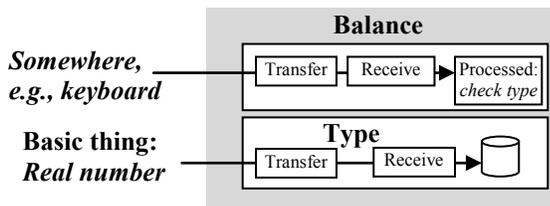

Fig. 26. Example of declaring a *type*

## VI. CONCLUSION

This paper has proposed a conceptual model, called the *Flow(thing) Machines* (FM) model, that produces a detailed class description using a single diagrammatic language. The result is an abstract picture (i.e., with no technical notions, e.g., software) that describes the total system for the purpose of understanding, design, and documentation. In the examples, such an idea is applied at the single class level, a class with simple (one level) inheritance, in addition to several simple relationships.

The FM methodology is founded on the machines mentioned above arranged into spheres and sub-spheres. The initial indication from the examples is that a class is a type of sphere. Relationships among classes are represented by flows and the overlapping of spheres. Class in FM does not have a structural character, but also, metaphorically speaking, it is a map of a city that comes alive with flows and events that form another dimension for describing the city. Continuing metaphorically, a tourist has a map of the city and also a description of events such as a trip that starts at a certain time in a certain place, while traveling on a tourist bus that flows from one place to another, with the first stop to eat at a certain place, … etc. This analogy resembles *methods*. Methods are pre-recognized events that occur over regions of the "map of the city." The interesting thing in this approach is that "other diagrams" (e.g., activity, state, sequence) have "crept back" into the class diagram, all dressed in the FM diagrammatic language.

Of course such an idea is still very elementary in comparison with the UML industrial marvel. Is the FM model as described in this paper a domain model that reflects the UML technical model? If it is, then it brings domain experts into the arena of class diagrams. Or, the FM model could be the job of domain experts utilized to communicate domain requirements to a development team that converts them to a more technical specification such as UML diagrams.

Regardless of the usability of FM in domain modeling or technical design, the general conclusion is that the object-oriented notions of diagrams are still not the last word in the field, and further approaches such as OPM [9] and FM may lead to development of a unified modeling methodology.

Future work will experiment with the various notions of the object-oriented paradigm, especially notions other than structure and behavior.

REFERENCES

[1]  J. Joque, "The invention of the object: object orientation and the philosophical development of programming languages", *Philosophy & Technology*, vol. 29, no. 4, pp. 335–356, 2016.

[2]  L. A. Maciaszek, *Requirements Analysis and System Design*, 3rd ed., Addison Wesley, 2007.

[3]  H. Washizaki, M. Akimoto, A. Hasebe, A. Kubo, and Y. Fukazawa, "TCD: a text-based UML class diagram notation and its model converters," in *Advances in Software Engineering, International Conference, ASEA 2010*, Springer, pp. 296–302, 2010.

[4]  D. Rayside and G. T. Campbell, "An Aristotelian understanding of object-oriented programming," *Conference on Object-Oriented Programming, Systems, Languages, and Applications*, October 15-19, Minneapolis, MN, USA, 2000.

[5]  L. Mota, L. Botelho, H. Mendes, and A. Lopes, "O3F: an object oriented ontology framework," *AAMAS'2003, 2nd International Joint Conference on Autonomous Agents and Multi-Agent Systems*, July 14–18, Melbourne, Austrália, 2003.

[6]  H. Eichelberger, "Nice class diagrams admit good design?" in *SoftVis '03: Proceedings of the 2003 ACM Symposium on Software Visualization*, pp. 159–168. ACM Press, 2003.

[7]  D. Sun and K. Wong, "On evaluating the layout of UML class diagrams for program comprehension," in *Proc. 13th IEEE Int. Workshop on Program Comprehension*, St. Louis, MO., USA, pp. 317–328, 2005.

[8]  T. Dwyer, "Three dimensional UML using force directed layout," in *CRPITS '01: Australian Symposium on Information Visualisation*, pp. 77–85, Australian Computer Society, 2001.

[9]  D. Dori, *Object-Process Methodology: A Holistic Systems Paradigm* Berlin: Springer,, 2002.

[10]  D. Dori, Modeling Knowledge with Object-Process Methodology, 2011 [Online]. http://esml.iem.technion.ac.il/wp-content/uploads/2011/08/Object-Process-Methodology.pdf

[11]  S. Al-Fedaghi, "Modeling events as machines," *Int. J. Comput. Sci. Inform. Sec.*, vol. 15, April 2017.

[12]  S. Al-Fedaghi, "Designing home circulation: application to smart homes," *Int. J. Smart Home*, vol. 10, no. 12, 2016.

[13]  S. Al-Fedaghi, "Philosophy made (partially) structured for computer scientists and engineers, *Int. J. u- and e- Service, Sci. Tech.*, vol. 9, no. 8, 2016.

[14]  S. Al-Fedaghi, "Flow-based conceptual representation of problems," *J. Theor. Appl. Inform. Technol.*, vol. 58 no. 2, pp. 347–356, 2013.

[15]  S. Al-Fedaghi, "Conceptual modeling in simulation: a representation that assimilates events," *Int. J. Adv. Comput. Sci. Appl.*, vol. 7, no. 10, pp. 281–289, 2016.

[16]  S. Al-Fedaghi, "Business process modeling: blueprinting," *Int. J. Comput. Sci. Inform. Security*, vol. 15, no. 3, pp. 286-291, 2017.

[17]  H. M. Deitel and P. J. Deitel, *C++ How to Program*, 5th ed. Prentice Hall, 2005. ISBN-10 0-13-185757-6

[18]  Wikibooks, Programming Concepts: Object-oriented programming (OOP), Last edited on 8 January 2016. https://en.wikibooks.org/wiki/A-level_Computing_2009/AQA/Problem_Solving,_Programming,_Operating_Systems,_Databases_and_Networking/Programming_Concepts/Object-oriented_programming_(OOP)#Classes

[19]  C. Hock-Chuan, Programming notes, C++ Programming Language, Object-Oriented Programming (OOP) in C++, Last modified: May 2013, https://www3.ntu.edu.sg/home/ehchua/programming/cpp/cp3_OOP.html

[20]  Source Code Solutions, UML diagrams for ATM (Automated Teller Machine) System, accessed June 2017. http://www.sourcecodesolutions.in/

[21]  L. de-Marcos, F. Flores, and J. J. Martínez, "Modeling with Plato: the Unified Modeling Language in a cultural context," *15th Annual Conference on Innovation and Technology in Computer Science (ITiCSE'10)*, Bikent, Ankara, Turkey, 2010.

[22]  M. A. Bunge, *Ontology II: A World of Systems*, vol. 4 of *Treatise On Basic Philosophy*. Dordrecht, Netherlands: D. Reidel, 1979.

[23]  J. M. Evermann, Using Design Languages for Conceptual Modelling: The UML Case, Ph.D. thesis, Wirtschafts-Diplom Informatik [Business Science Degree, IT], Westfälische Wilhelms-Universität Münster, 1998.